\documentclass[twocolumn,amsmath,amssymb,prb,floatfix,superscriptaddress,showpacs]{revtex4-1}
\usepackage{graphicx}
\usepackage{dcolumn}
\usepackage{bm}
\usepackage[flushleft]{threeparttable}
\usepackage{xspace}
\usepackage{hyperref}
\usepackage{color}
\usepackage{multirow}
\def\new{\color{black}}

\def\rucl{$\alpha$-RuCl$_3$\xspace}
\providecommand*{\et}{\emph{et\,al.}\xspace}%
\def\yzgo{YbZnGaO$_4$\xspace}
\def\ymgo{YbMgGaO$_4$\xspace}
\begin{document}

\title{Experimental identification of quantum spin liquids}
\author{Jinsheng~Wen}
\email{jwen@nju.edu.cn}
\author{Shun-Li~Yu}
\affiliation{National Laboratory of Solid State Microstructures and Department of Physics, Nanjing University, Nanjing 210093, China}
\affiliation{Collaborative Innovation Center of Advanced Microstructures, Nanjing University, Nanjing 210093, China}
\author{Shiyan Li}
\affiliation{State Key Laboratory of Surface Physics, Department of Physics, and Laboratory of Advanced Materials, Fudan University, Shanghai 200433, China}
\affiliation{Collaborative Innovation Center of Advanced Microstructures, Nanjing University, Nanjing 210093, China}
\author{Weiqiang Yu}
\affiliation{Department of Physics and Beijing Key Laboratory of
Opto-electronic Functional Materials $\&$ Micro-nano Devices, Renmin
University of China, Beijing, 100872, China}
\author{Jian-Xin Li}
\affiliation{National Laboratory of Solid State Microstructures and Department of Physics, Nanjing University, Nanjing 210093, China}
\affiliation{Collaborative Innovation Center of Advanced Microstructures, Nanjing University, Nanjing 210093, China}

\begin{abstract}
{\bf In condensed matter physics, there is a novel phase termed ``quantum spin liquid", in which strong quantum fluctuations prevent the long-range magnetic order from being established, and so the electron spins do not form an ordered pattern but remain ``liquid" like even at absolute zero temperature. Such a phase is not involved with any spontaneous symmetry breaking and local order parameter, and to understand it is beyond the conventional phase transition theory. Due to the rich physics and exotic properties of quantum spin liquids, such as the long-range entanglement and fractional quantum excitations, which are believed to hold great potentials in quantum communication and computation, they have been intensively studied since the concept was proposed in 1973 by P.~W.~Anderson. Currently, experimental identifications of a quantum spin liquid still remain as a great challenge. Here, we highlight some interesting experimental progress that has been made recently. We also discuss some outstanding issues and raise questions that we consider to be important for future research.}
\end{abstract}

\maketitle

\section{The road to quantum spin liquids}
\begin{figure*}[htb]
\centering
\includegraphics[width=0.98\linewidth,trim=0mm 0mm 0mm 0mm]{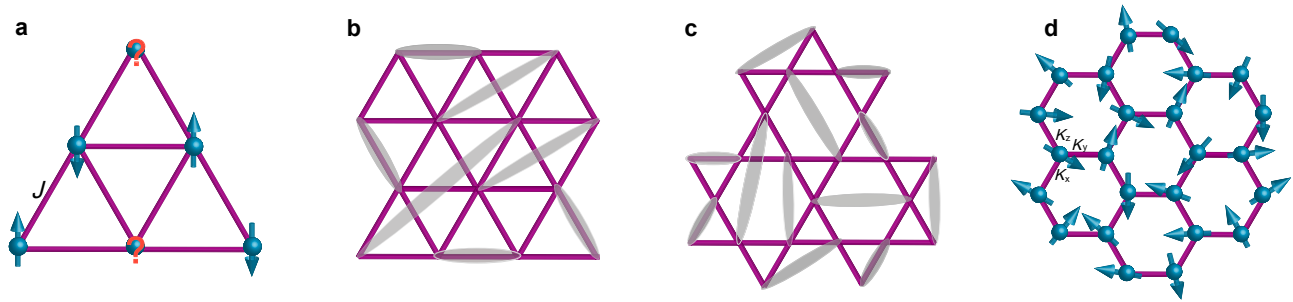}\\[5pt]
\caption{\label{qslstructures}{{\bf Schematics of some typical two-dimensional crystal structures on which quantum spin liquids may be realized.} {\bf a} and {\bf b} Triangular, {\bf c} Kagome, and {\bf d} Honeycomb structures. Arrows and question marks represent spins and undetermined spin configurations due to geometrical frustration, respectively. $J$ is the Heisenberg exchange interaction, and $K_{x,y,z}$ are Kitaev interactions along three bonds. The shades illustrate spin singlets with a total spin $S=0$, each formed by two spins antiparrallel to each other.}}
\end{figure*}

As we learn from the textbook, a magnetic material tends to have their electron spins arrange into a certain, for example, parallel or antiparallel pattern at low temperatures to minimize the free energy. The two patterns correspond to the ferromagnet and antiferromagnet, which are governed by a negative and positive superexchange coupling constant $J$, respectively. However, for materials that have competing magnetic interactions, the situation sometimes can be complicated. For example, in each of the edge-shared triangular lattice as sketched in Fig.~\ref{qslstructures}a, a third spin cannot align antiparallelly to the other two to satisfy the $J>0$ condition simultaneously. As such, the magnetic interaction of this spin is frustrated due to the triangular geometry, and the spin direction remains undetermined.\cite{nature464_199} Later on, it has been shown that there is a classical solution to avoid such a geometrically frustrated configuration on the triangular lattice---each spin on a triangle points to 120$^\circ$ with respect to each other, leading to a ground state with the 120$^\circ$ long-range magnetic order.\cite{PhysRevB.50.10048}

In 1973, P.~W.~Anderson proposed a more exotic ground state to describe the antiferromagnetically interacting spins on the two-dimensional triangular lattice.\cite{Anderson1973153} The model used to describe this state is the resonating-valence-bond (RVB) model, in which any two antiparallel spins pair up to form a spin singlet, with a total spin $S=0$ and vanishing net magnetic moment. Due to strong quantum fluctuations arising from the geometrical frustration, there is no particular arrangement of these singlets. In a technical language, the wavefunction of the RVB state is a linear superposition of all possible configurations of the singlets, as illustrated in Fig.~\ref{qslstructures}b. In this case, the strongly interacting spins do not develop a long-range magnetic order even at the absolute zero temperature, and the spin pattern for such a state is like a liquid. Such a novel quantum state is termed the quantum spin liquid (QSL), which has the following features: absence of long-range magnetic order, no spontaneous symmetry breaking of the crystal lattice or spins, long-range entanglement between the spins and as a consequence, fractional spin excitations. Note that a QSL is fundamentally distinct from a classical spin liquid. The latter is because the thermal fluctuation energy $k_{\rm B}T$ dominates over $J$, and the magnetic order is not established at high temperatures. The defining features---long-range entanglement and fractional spin excitations for a QSL are absent in such a classical state. {\new The physics is driven by the spin dynamics, instead of the statics, which shows interesting effects from spin-spin correlation that  are quantum in nature at low temperatures ($i.e., k_{\rm B}T \ll J$).} The RVB proposal initiated enormous efforts seeking for QSLs both on the theoretical and experimental sides.\cite{0143-0807-21-6-302,nature464_199,0034-4885-80-1-016502,RevModPhys.89.025003}.

Compared to the triangular lattice, quantum fluctuations due to the geometrical frustration and low coordinations are even stronger in the kagome lattice, which has six corner-shared triangles surrounding a hexagon as illustrated in Fig.~\ref{qslstructures}c. For the kagome lattice, geometrical frustration cannot be bypassed by having the 120$^\circ$ order as does the triangular lattice.\cite{PhysRevB.50.10048} Theoretical calculations suggest that the ground state for the kagome lattice is a QSL, although the detailed classification for such a state is still under hot debate.\cite{PhysRevB.45.12377,Yan1173,PhysRevB.95.235107,PhysRevLett.98.117205,PhysRevLett.118.137202}
We should also mention that in the pyrochlore lattice which hosts the spin-ice phase, there also exists large geometrical frustration, and the QSL phase is also possible in principle, in the case of small spins.\cite{nature464_199,Bramwell1495}

QSLs resulting from geometrical frustration in the triangular, kagome and pyrochlore lattices are usually inferred from approximations and conjectures, as the quantum spin model with frustrations can be hardly solved at present. On the other hand, in 2006, Alexei Kitaev\cite{aop321_2} proposed an exactly-solvable $S=1/2$ model on the honeycomb lattice (Fig.~\ref{qslstructures}d), which has substantially advanced the field. This model, as in Eq.~\ref{kitaevmodel}, is named the Kitaev model.
\begin{equation}
\label{kitaevmodel}
H=-K_x\sum_{x\text{-bonds}}S_i^xS_j^x-K_y\sum_{y\text{-bonds}}S_i^yS_j^y-K_z\sum_{z\text{-bonds}}S_i^zS_j^z
\end{equation}
Here, $H$ is the Hamiltonian, and $K_{x,y,z}$ are the nearest-neighbour Ising-type spin interactions along the $x$, $y$, and $z$ bonds (Kitaev interactions). It has been shown that the exact ground state of the Kitaev model is a QSL, which may be gapped or gapless depending on the relative strength of the Kitaev interactions along the three bonds.\cite{aop321_2} Unlike QSLs described by the RVB model, the QSL state in the Kitaev model (Kitaev QSL hereafter) is defined on the honeycomb lattice where geometrical frustration is absent. Instead, the presence of bond-dependent Kitaev interactions induces strong quantum fluctuations that frustrate spin configurations on a single site, resulting in a Kitaev QSL state.

The Kitaev model is simple and beautiful but was considered as a ``toy" model in the beginning, since the underlying anisotropic Kitaev interactions are unrealistic for a spin-only system. One brilliant proposal to realize the Kitaev interaction is to include the orbital degree of freedom in a Mott insulator that arises from the combination of the crystal electric field, spin-orbital coupling (SOC), and electron correlation effects.\cite{prl102_017205} A Mott insulator is an insulator which band theory predicts to be metallic but turns out to be insulating due to electron-electron interactions.\cite{lee:17} In the proposal,\cite{prl102_017205} the electronic configuration is a consequence of the following processes: First, the octahedral crystal field splits the $d$ orbitals into triply degenerate $t_{2g}$ and doubly degenerate $e_g$ orbitals at low and high energies, respectively; Second, in the presence of strong SOC, the degeneracy of the $t_{2g}$ is lifted, giving rise to one effective spin $J_{\rm eff}=1/2$ and one $J_{\rm eff}=3/2$ band, both of which are entanglements of the $S=1/2$ spin and $L_{\rm eff}=-1$ orbital moment for $t_{2g}$; Third, the resulting $J_{\rm eff}=1/2$ band is narrower than the original triply degenerate $t_{2g}$ band. Consequently, a moderate electron correlation will open a gap and split this band into a lower and upper Hubbard band. Such a state is a $J_{\rm eff}=1/2$ Mott insulator. Due to the spatial anisotropy of the $d$ orbitals and strong SOC, the exchange couplings between the effective spins are intrinsically anisotropic and bond dependent, and will give rise to the Kitaev interactions under certain circumstance.\cite{aop321_2,prl102_017205} These advancements bring the hope to find Kitaev QSLs in real materials, and have motivated intensive research in searching for QSLs beyond the triangular, kagome and pyrochlore lattices in recent years.\cite{doi:10.1146/annurev-conmatphys-033117-053934,0953-8984-29-49-493002}

\section{Why quantum spin liquids are of interest}

QSL is a highly nontrivial state of matter. Ordinary states are often understood within Landau's framework in terms of order, phase transition, and symmetry breaking. However, a QSL does not have a local order parameter and avoids any spontaneous symmetry breaking even in the zero-temperature limit, so to understand it requires theories beyond Landau's phase transition theory. In fact, most QSLs are believed to be associated with topological order and phase transition.\cite{np6_376,RevModPhys.89.041004} 

QSLs are Mott insulators where the bands are half filled, different from band insulators where the bands are fully filled or empty. Here, the electron-electron correlation is important, and the conventional band theory with the single-particle approximation does not work well.\cite{arcmp5_57,arcmp7_195} Therefore, understanding QSLs should and has already provided deep insights into the rich physics of strongly correlated electron systems. Moreover, in 1987, a year after high-temperature superconductivity was discovered in copper oxides,\cite{mullerlbco} P.~W.~Anderson proposed that the parent compound, a Mott insulator La$_2$CuO$_4$, which by doping with carriers can superconductivity be achieved, was a QSL.\cite{anderson1} This proposal injects significant momentum into the research of QSLs as it may lend support to solve the puzzle of high-temperature superconductivity.\cite{lee:17}

Besides the fascinating physics associated with QSLs, they also hold great application potentials. For example, in a QSL, the spins are entangled over a long range, which is an essential ingredient for quantum communication.\cite{np8_902} Another important property of a QSL associated with the long-range entanglement is the presence of fractional spin excitations. For instance, a Kitaev QSL can support fractional excitations represented by Majorana fermions, {\new which can be made to behave as anyons obeying the non-Abelien statistics in the presence of magnetic field.\cite{aop321_2}} Braiding these anyons is an important step towards the topological quantum computation.\cite{Kitaev20032}

\begin{table*}[htb]
  \begin{threeparttable}
\caption{Geometrically-frustrated quantum-spin-liquid candidates}
\label{tbl:geo}
\begin{tabular*}{\textwidth}{@{\extracolsep{\fill}}cccccccc}
\hline \hline
Material & Structure & $\Theta_{\rm CW}$ (K) & $J$ (K) & Reference \\
\hline
$\kappa$-(BEDT-TTF)$_2$Cu$_2$(CN)$_3$ & Triangular  & -375 & 250  & \onlinecite{PhysRevLett.91.107001} \\
EtMe$_3$Sb[Pd(dmit)$_2$]$_2$ & Triangular  & -(375-330) & 220-250 & \onlinecite{np6_673} \\
\ymgo & Triangular & -4 & 1.5 & \onlinecite{sr5_16419} \\
1T-TaS$_2$ & Triangular & -2.1 & 0.1 & \onlinecite{PhysRevB.96.195131}\\
ZnCu$_3$(OH)$_6$Cl$_2$ (herbertsmithite) & Kagome & -314 & 170 & \onlinecite{doi:10.1021/ja053891p} \\
Cu$_3$Zn(OH)$_6$FBr (barlowite) & Kagome & -200 & 170 & \onlinecite{ZiliFeng:77502}\\
{\new Na$_4$Ir$_3$O$_8$} & {\new Hyperkagome } & {\new -650 } & {\new 430 }  & {\new \onlinecite{prl99_137207} } \\
PbCuTe$_2$O$_6$ & Hyperkagome & -22 & 15 & \onlinecite{prb90_035141} \\
Ca$_{10}$Cr$_7$O$_{28}$ & Distorted kagome &  & -9 & \onlinecite{np12_942} \\
\hline \hline
\end{tabular*}
\begin{tablenotes}
\item $\Theta_{\rm CW}$ and $J$ are the Curie-Weiss temperature and exchange coupling constant, respectively.\\
\vspace{2mm}
$^*$Notes:\\
\quad $\kappa$-(BEDT-TTF)$_2$Cu$_2$(CN)$_3$: With a $J\sim250~K$, no magnetic order nor spin freezing is observed down to 20~mK,\cite{PhysRevB.73.140407} giving rise to a frustration index over 10000. While low-temperature specific heat measurements indicate the presence of gapless magnetic excitations,\cite{np4_459} they do not contribute to the thermal conductivity.\cite{np5_44}\\
\quad \ymgo: A frequency-dependent peak in the a.c. susceptibility characteristic of a spin glass is observed around 0.1~K.\cite{PhysRevLett.120.087201}\\
\quad 1T-TaS$_2$: Note that $\Theta_{\rm CW}$ and $J$ listed in the table are significantly smaller than those calculated from the high-temperature susceptibility in ref.~\onlinecite{np13_1130}. In 1T-TaS$_2$, 13 Ta$^{4+}$ ions form a hexagonal David-star cluster when the system is cooled below the commensurate-charge-density-wave phase transition temperature at 180~K.\cite{Law03072017} Twelve Ta$^{4+}$ ions in the corner of the star form 6 covalent bonds and the spins connected by the bonds cancel out. An orphan electron is localized in the centre of the David star, giving rise to a net spin of $S=1/2$ for each David star. The spins in the centre constitute a triangular network, making the system essentially a geometrically frustrated magnet. At present, whether or not this material is a QSL is still under hot debate.\cite{Law03072017,npjqm2_42,np13_1130,PhysRevB.96.081111,PhysRevB.96.195131}\\
\quad {\new Na$_4$Ir$_3$O$_8$: Although a well-defined magnetic phase transition is not observed in Na$_4$Ir$_3$O$_8$, a spin-glass transition occurs around 6~K, below which the spins are frozen and maintain short-range correlations.\cite{prl99_137207,PhysRevLett.113.247601,PhysRevLett.115.047201} } \\
\quad Ca$_{10}$Cr$_7$O$_{28}$: It is a system with complex structure, and more interestingly, with dominant ferromagnetic interactions.\cite{np12_942} Frequency-dependent peak was observed in the a.c. susceptibility, suggestive of a spin-glass ground state, but it is suggested that such a state could be ruled out by performing the Cole-Cole analysis on the a.c. susceptibility data.\cite{np12_942}
\end{tablenotes}
\end{threeparttable}
\end{table*}

\begin{table*}[htb]
\caption{Kitaev quantum-spin-liquid candidates}
\label{tbl:kitaev}
\begin{tabular*}{\textwidth}{@{\extracolsep{\fill}}cccccccc}
\hline \hline
Material & Crystal structure & Dimensionality & $T_N$ (K) & Magnetic structure & Moment & Reference \\
\hline
Na$_2$IrO$_3$ & Honeycomb & 2 & 13-18  & Zigzag & 0.22~$\mu_{\rm B}$/Ir$^{4+}$  & \onlinecite{PhysRevB.82.064412} \\
$\alpha$-Li$_2$IrO$_3$ & Honeycomb & 2 & 15 & Incommensurate & 0.4~$\mu_{\rm B}$/Ir$^{4+}$ & \onlinecite{PhysRevLett.108.127203} \\
$\beta$-Li$_2$IrO$_3$ & Hyperhoneycomb & 3 & 37 & Incommensurate & 0.47~$\mu_{\rm B}$/Ir$^{4+}$ & \onlinecite{PhysRevB.90.205116} \\
$\gamma$-Li$_2$IrO$_3$ & Stripyhoneycomb & 3 & 39.5 & Incommensurate &  & \onlinecite{PhysRevLett.113.197201} \\
$\alpha$-RuCl$_3$ & Honeycomb & 2 & 8 & Zigzag & 0.4~$\mu_{\rm B}$/Ru$^{3+}$ & \onlinecite{PhysRevB.91.144420} \\
\hline \hline
\end{tabular*}
\end{table*}

\section{Candidate materials}
Motivated by the rich physics and fascinating properties of QSLs, studying these materials has been a 45-year long but still dynamic field. There has been a lot of progress made both on the theoretical and experimental aspects. QSLs in one-dimensional magnetic chain have been well established and will not be discussed in this work. For reference, there is a recent review article in this topic in ref.~\onlinecite{npjqm3_18}. Here, we will focus on QSLs in two or three dimensions. There have been many theoretical proposals for QSLs. For details, please refer to the reviews and references therein.\cite{0034-4885-80-1-016502,RevModPhys.89.025003,0953-8984-29-49-493002}. 

Experimentally, numerous materials have been suggested to be candidates for QSLs, some of which are tabulated in Tables~\ref{tbl:geo} and \ref{tbl:kitaev}. We classify these materials into two categories--- geometrically frustrated materials where the RVB model is applicable (Table~\ref{tbl:geo}) and Kitaev QSL candidates where the Kitaev physics is relevant (Table~\ref{tbl:kitaev}). For the majority of these materials, the discussions of QSL physics are in two dimensions, where the reduced dimensionality enhances quantum fluctuations needed for QSLs. In recent years, there have also been attempts to look for three-dimensional QSL candidates, such as hyperkagome Na$_4$Ir$_3$O$_8$~(ref.~\onlinecite{prl99_137207}) and  PbCuTe$_2$O$_6$,\cite{prb90_035141} hyperhoneycomb $\beta$-Li$_2$IrO$_3$, \cite{PhysRevB.90.205116} and stripyhoneycomb $\gamma$-Li$_2$IrO$_3$.\cite{PhysRevLett.113.197201} A more detailed review on the search of three-dimensional QSLs can be found in ref.~\onlinecite{0953-8984-29-49-493002}.

\section{How to identify a quantum spin liquid experimentally}

\subsection{Technique overview}

Strictly speaking, to identify a QSL, one should perform measurements at absolute zero temperature, which is not practical. As a common practice, it is often assumed that when the measuring temperature $T$ is far below the temperature which characterizes the strength of the magnetic exchange coupling, say two orders of magnitude for instance, the finite-temperature properties can be a representation of those of the zero-temperature state, provided that there is no phase transition below the measuring temperature. Even so, experimental identification of a QSL is still challenging since a QSL does not spontaneously break any lattice symmetry and no local order parameter exists to describe such a state. One first step is to prove that there is no magnetic order nor spin freezing down to the lowest achievable temperature in, preferably, low-spin systems whose spin interactions are dominantly antiferromagnetic. For example, measuring the magnetic susceptibility is very useful in this aspect. {\new By performing a Curie-Weiss fit to the high-temperature magnetic susceptibility,} one can also obtain the Curie-Weiss temperature $\Theta_{\rm CW}$ that characterizes the strength of the antiferromagnetic interactions. We list the $\Theta_{\rm CW}$ and exchange coupling constant $J$ for some materials in Table~\ref{tbl:geo}.

Specific heat is also quite often measured to help screen a QSL. For a material that orders magnetically at low temperatures, a $\lambda$-type peak in the specific heat {\it vs} temperature curve is generally expected. By contrast, for a QSL, such a sharp peak is typically absent, unless in the case such as  that there is a topological phase transition.\cite{PhysRevLett.113.197205} Furthermore, it also reveals that whether there exist finite magnetic excitations at low temperatures or not. The residual magnetic entropy can be calculated, provided that the phonon contribution to the specific heat can be properly subtracted, ideally by subtracting the specific heat of a proper nonmagnetic reference sample. By examining how much entropy has already been released at the measuring temperature, the possibility of establishing a long-range magnetic order at a lower temperature can be estimated.\cite{sr5_16419,PhysRevLett.120.087201} In addition to these macroscopic measurements, microscopic probes such as muon-spin-relaxation ($\mu$SR) and nuclear magnetic resonance (NMR) that are sensitive to the local magnetic environment are also often used to detect whether there is spin ordering or freezing.\cite{np12_942,np13_1130,prl115_167203} Neutron diffraction can also tell explicitly whether a system is magnetically ordered or not.

These measurements reveal very important information but are not completely satisfactory, partly because: even though magnetic ordering can be excluded, {\new such a ground state without a long-range order may be caused by the disorder which leads to a percolative breakdown of the long-range order.} In such a case, the spin-liquid-like state is similar to a classical spin liquid, but essentially different from a QSL. So, are there better approaches that can identify a QSL more directly from a positive aspect?

The key features defining a QSL are the the long-range entanglement and the associated fractional spin excitations. At present, it is hard to define and characterize the former, so the answer to this question mostly lies on the fractional excitations, {\it e.g.}, spinons predicted in the RVB model.\cite{Anderson1973153} {\new Spinons are charge-neutral fermionic quasiparticles carrying fractional spins such as $S=1/2$ resulting from the fractionalization of the $S=1$ excitations.} They are deconfined from the crystal lattice, have their own dispersions, and become itinerant in the crystal. For certain type of QSLs, for instance, U(1) gapless QSLs, in which the low-energy effective model has a U(1) gauge structure, spinons form a Fermi sea, similar to that formed by electrons in a metal.\cite{PhysRevB.70.214437} Techniques that are sensitive to magnetic excitations such as inelastic neutron scattering (INS), thermal conductivity and thermal Hall conductivity, NMR, electron-spin resonance (ESR), specific heat, as well as Raman and terahertz (THz) spectroscopies, can be utilized to clarify whether fractional excitations are present in a QSL candidate, and to classify the type of a particular QSL based on the behaviours of the fractional excitations. Below, we provide some example studies on QSL candidates using these techniques.

\subsection{Examples}

\begin{figure}[htb]
\centerline{\includegraphics[width=\linewidth]{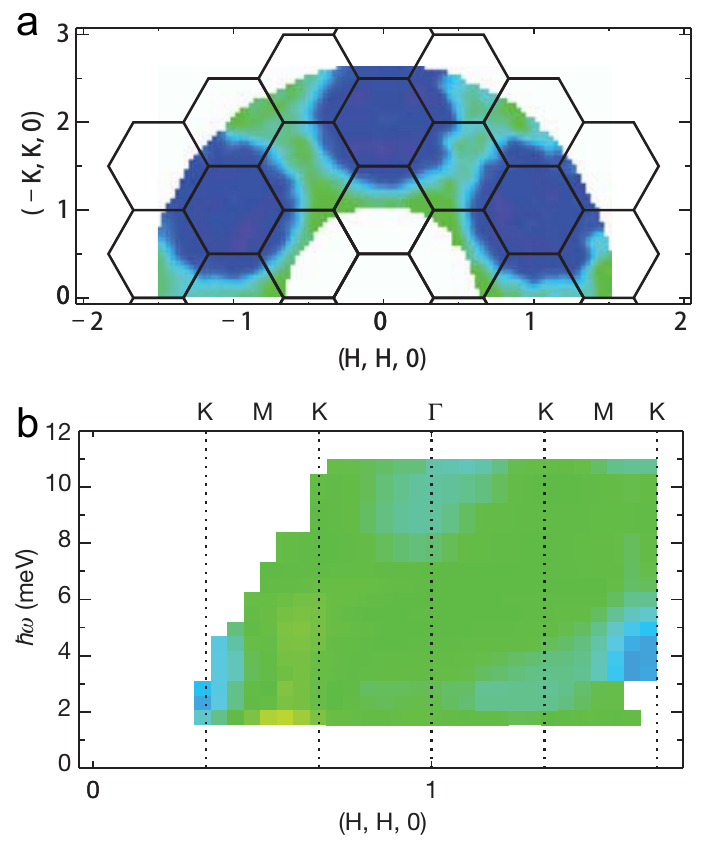}}
\caption{{\bf Broad continuous magnetic excitation spectra of herbertsmithite, ZnCu$_3$(OH)$_6$Cl$_2$.} {\bf a} Constant-energy false-colour contour plot of the spin excitation spectra obtained from INS measurements. Solid lines indicate the Brillouin zone boundaries. {\bf b} Magnetic dispersion along a high-symmetry direction [110]. Reprinted with permission from ref.~\onlinecite{nature492_406}.}
\label{herbertsmithitecontinuum}
\end{figure}

One of the most convincing evidences for the presence of deconfined spinons with fractional quantum number $S=1/2$ is the observation of broad continuum magnetic excitation spectra in INS measurements. Since magnetic neutron scattering is a spin-1 process, at least one spinon pair must be excited in a spin-flip event. In this case, any pair of the spinons that satisfies the energy-momentum conservation rule $E_{q}=\epsilon_s(k)+\epsilon_s(q-k)$ can be excited. Here, $\epsilon_s(k)$ is the dispersion of spinons, and $E$ and $q$ are the energy and momentum transfers, respectively. Since there are many $k$ values that obey this rule for any particular combination of $E$ and $q$, a continuous spectrum which is broad both in momentum and energy can be observed by INS measurements for a QSL whose fractional excitations are spinons. Such a continuum has been reported in various compounds, such as ZnCu$_3$(OH)$_6$Cl$_2$ (herbertsmithite),\cite{nature492_406} Ca$_{10}$Cr$_7$O$_{28}$,\cite{np12_942} Ba$_3$NiSb$_2$O$_9$,\cite{PhysRevB.95.060402} and \ymgo.\cite{nature540_559,np13_117} An example of the broad continuum in herbertsmithite is shown in Fig.~\ref{herbertsmithitecontinuum}.\cite{nature492_406} The spectra in the momentum space in Fig.~\ref{herbertsmithitecontinuum}a are diffusive and elongating along the $\langle100\rangle$ directions. The dispersions in Fig.~\ref{herbertsmithitecontinuum}b are continuous in the whole energy range measured. These results are distinct from well-defined spin excitations in the magnetic order state, which are sharply peaked around the ordering wavevector. Since different types of QSLs will give rise to different excitation spectra,\cite{arXiv:1507.03007} analyzing the detailed momentum and energy dependence of the continuous spectra will help classify the QSLs. For instance, for ZnCu$_3$(OH)$_6$Cl$_2$, based on the spectral weight distribution in the momentum space,\cite{nature492_406} it is suggested that this compound is a Dirac QSL, featuring linearly dispersing spinonon bands.\cite{arXiv:1507.03007} On the other hand, Ba$_3$NiSb$_2$O$_9$ and \ymgo have been classified to be the U(1) QSL having a large spinon Fermi surface.\cite{PhysRevB.95.060402,nature540_559} 

\begin{figure}[htb]
\centerline{\includegraphics[width=0.98\linewidth]{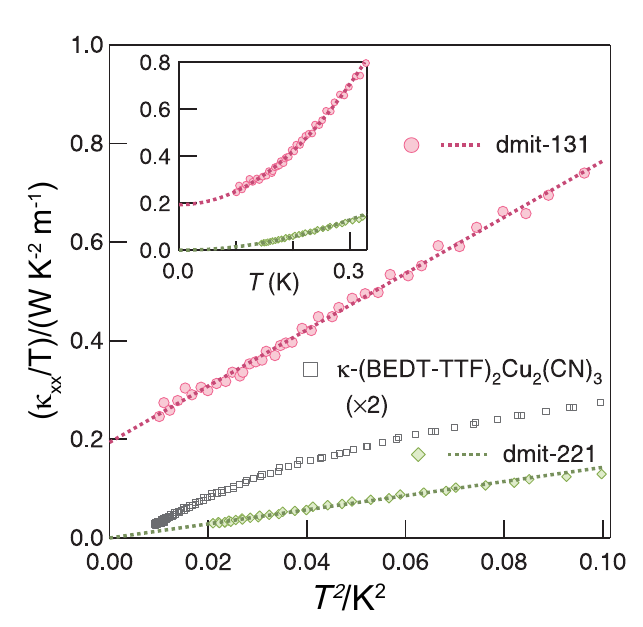}}
\caption{{\bf Thermal conductivity of EtMe$_3$Sb[Pd(dmit)$_2$]$_2$.} Thermal conductivity $\kappa_{xx}$ over temperature $T$ as a function of $T^2$ at low temperatures for EtMe$_3$Sb[Pd(dmit)$_2$]$_2$ (dmit-131) (circles), Et$_2$Me$_2$Sb[Pd(dmit)$_2$]$_2$ (dmit-221) (diamonds), and $\kappa$-(ET)$_2$Cu$_2$(CN)$_3$ (squares). The inset shows $\kappa_{xx}/T$ plotted as a function of $T$. Reprinted with permission from ref.~\onlinecite{Yamashita1246}.}
\label{3-2}
\end{figure}

In the mean-field level, spinons are free fermions. So if spinons exist  at low energies, the spinon term ($\kappa_{\rm s}$) should dominate the thermal conductivity, $\kappa$, as the electron ($\kappa_{\rm e}$) and phonon terms ($\kappa_{\rm p}$) in $\kappa=\kappa_{\rm e}+\kappa_{\rm p}+\kappa_{\rm s}$ are either zero or small at low temperatures. In Fig.~\ref{3-2}, the thermal conductivity over $T$ ($\kappa_{xx}/T$) of EtMe$_3$Sb[Pd(dmit)$_2$]$_2$, a QSL candidate with the triangular-lattice structure, is plotted against $T^2$ in comparison with those of another triangular-lattice QSL candidate $\kappa$-(ET)$_2$Cu$_2$(CN)$_3$ and a nonmagnetic compound Et$_2$Me$_2$Sb[Pd(dmit)$_2$]$_2$.\cite{Yamashita1246} In contrast to the latter two, where $\kappa_{xx}/T$ is 0 as $T$ approaches 0, the residual $\kappa_{xx}/T$ for EtMe$_3$Sb[Pd(dmit)$_2$]$_2$ is 0.2~W\,K$^{-2}$\,m$^{-1}$.\cite{Yamashita1246} The presence of large residual $\kappa_{xx}/T$ is also confirmed in the inset of Fig.~\ref{3-2} in which $\kappa_{xx}/T$ is plotted as a function of $T$. This result is consistent with the observation of gapless magnetic excitations in the specific heat measurement.\cite{nc2_275} Empirically, $\kappa=(1/3)C_{\rm m}v_{\rm F}l$, where $C_{\rm m}$, $v_{\rm F}$, and $l$ are magnetic specific heat, Fermi velocity, and mean-free path, respectively. Using this formula, it is estimated the highly mobilized magnetic excitations have an $l$ of $\sim$1000 times of the spin-spin distance.\cite{Yamashita1246} These results indicate the presence of gapless magnetic excitations, which are a rare example where thermal conductivity measurements agree with the specific heat data in the studies of QSL candidates.\cite{nc2_275}

\begin{figure}[htb]
\centerline{\includegraphics[width=0.98\linewidth]{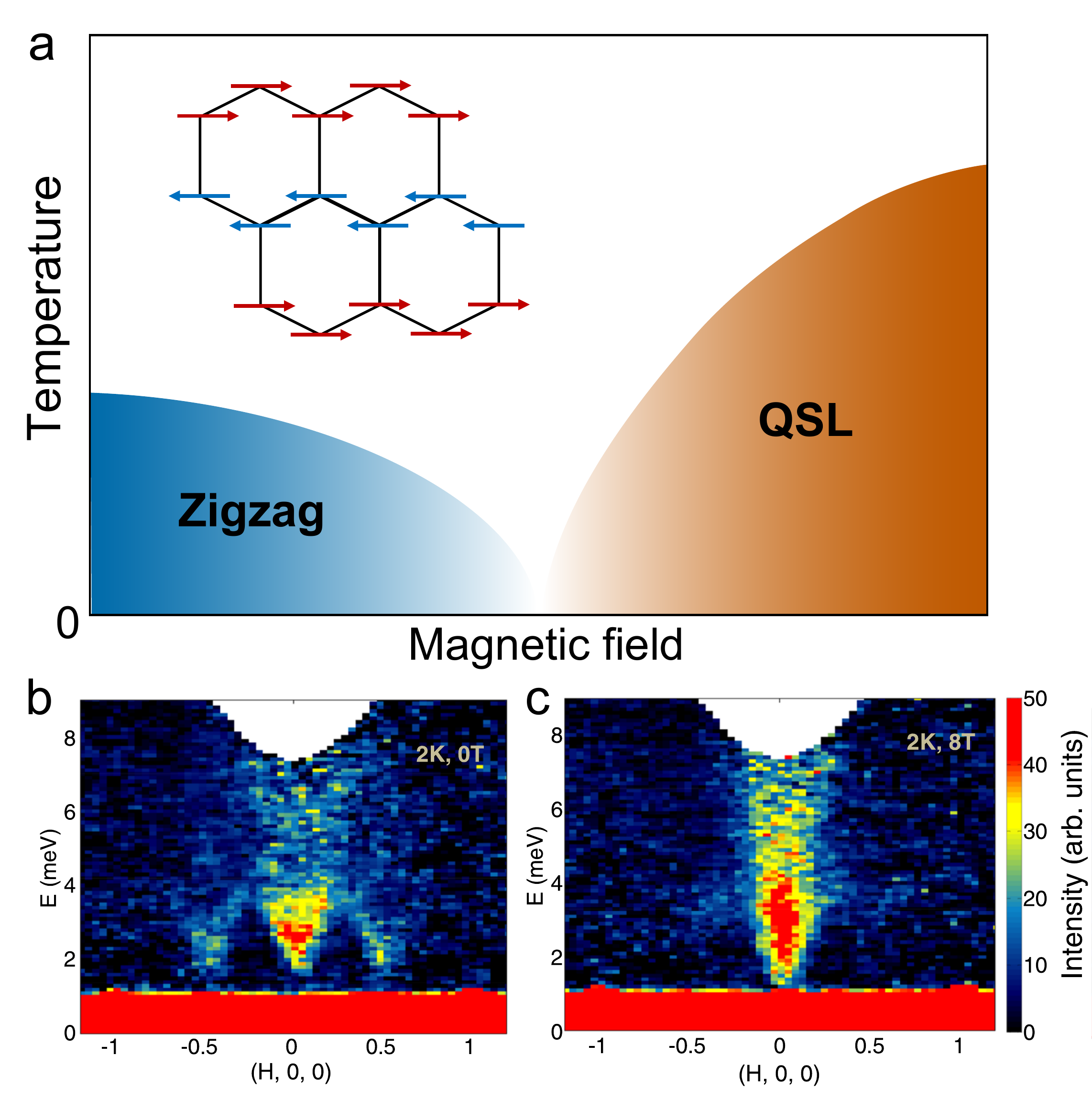}}
\caption{{\bf Phase diagram and magnetic excitations of \rucl.} {\bf a} A schematic phase diagram in the magnetic field and temperature space. The inset illustrates the zigzag magnetic order on the honeycomb lattice, with arrows representing spins. {\bf b} and {\bf c} Magnetic dispersions along the [100] direction at zero (zigzag order phase) and 8-T fields (QSL phase). {\bf b} and {\bf c} are reprinted with permission from ref.~\onlinecite{npjqm3_8}.}
\label{rucl}
\end{figure}

In recent years, $\alpha$-RuCl$_3$ with the honeycomb-lattice structure has been studied intensively for the sake of Kitaev QSLs. The reality is that the ground state of \rucl is not a Kitaev QSL, but a zigzag magnetic order state, as illustrated in the inset of Fig.~\ref{rucl}a.\cite{nature199_1089,PhysRevB.90.041112}
Starting from the five-orbital Hubbard model and based on the energy-band structure calculated from the first-principles method, Wang $et~al.$ have found that the minimal effective model to describe the zigzag order in $\alpha$-RuCl$_3$ is the so-called $K$-$\Gamma$ model,\cite{PhysRevB.96.115103,PhysRevLett.112.077204} where $K$ and $\Gamma$ represent the Kitaev and off-diagonal interactions, respectively. In particular, the Kitaev interaction is shown to be ferromagnetic and large in this material, demonstrating that the anisotropic Kitaev interaction underlying a Kitaev QSL can be realized in real materials.\cite{PhysRevLett.118.107203} {\new We note that there is still ongoing debate on the sign of the Kitaev interaction.\cite{0953-8984-29-49-493002,PhysRevB.96.064430,PhysRevB.97.241110}} Furthermore, INS results indicate that the zigzag ground state of \rucl may be proximate to the Kitaev QSL phase.\cite{nm15_733} As shown in Fig.~\ref{rucl}b, in the zigzag order phase, in addition to the spin-wave excitations dispersing from $(0.5,0,0)$ (the M point of the Brillouin zone), there also exist broad continuous excitations at the Brillouin zone centre ($\Gamma$ point), the latter of which are believed to be associated with fractionalized excitations resulting from the Kitaev QSL phase.\cite{Banerjee1055,np13_1079,np14_786}

The magnetic order is rather fragile, with an ordered moment of $\sim$0.4$\mu_B$ and an ordering temperature of $\sim$8~K.\cite{nature199_1089,PhysRevB.90.041112} Such a fragile order can be fully suppressed by either an in-plane magnetic field\cite{PhysRevB.91.094422} or pressure.\cite{PhysRevB.96.205147,PhysRevB.97.245149} Intensive research on the field effects in \rucl utilizing various techniques has resulted in many similar phase diagrams, which are summarized in the review article in ref.~\onlinecite{ZhenMa:106101}. In Fig.~\ref{rucl}a, we sketch an oversimplified phase diagram with only two phases, namely, the low-field zigzag order phase and the high-field disordered phase. It is now believed by many that above the critical field $B_c$ around 7.5~T, when the zigzag magnetic order is fully suppressed, the high-field disordered state is the QSL state, although the detailed nature of this phase is still under debate.\cite{ZhenMa:106101}

Banerjee $et~al$. have carried out INS measurements to examine the magnetic-field evolution of the magnetic excitations, and some of the results are shown in Fig.~\ref{rucl}b and c.\cite{npjqm3_8} They find that the spin-wave excitations associated with the zigzag order near the M point ({\new in the trigonal space group}) are suppressed with the field. As shown in Fig.~\ref{rucl}(c), above $B_c$, at $\mu_0H=8$~T, excitations near the M point are completely gone, and only broad continuous excitations believed to be resulting from the Kitaev QSL phase near the $\Gamma$ point remain. The high-field results are similar to the results under zero field above the ordering temperature.\cite{npjqm3_8} By comparing with calculations involving a pure Kitaev model, they suggest that the excitations under high fields resemble those of a Kitaev QSL.\cite{npjqm3_8} Very recently, thermal Hall conductivity measurements have observed half-integer quantized plateau around $B_c$, indicating that the phase in this regime is the topologically nontrivial chiral QSL, while the state for $B\gtrsim9$~T is not topological.\cite{nature559_227}

\section{Outstanding questions}

In recent years, a lot of progress has already been made in the research of QSLs, but for the existing QSL candidates, there are more or less some drawbacks and unresolved issues. For example, for the Kitaev QSL candidates listed in Table~\ref{tbl:kitaev}, the ground state exhibits either static magnetic order or short-range spin freezing.\cite{0953-8984-29-49-493002} For organic compounds such as $\kappa$-(BEDT-TTF)$_2$Cu$_2$(CN)$_3$, the specific heat indicates a gapless ground state,\cite{np4_459} but thermal conductivity measurements reveal no contributions from the magnetic excitations,\cite{np5_44} inconsistent with the gapless QSL state. Is the contradiction resulting from the disorder effect which makes the spinons localized and thus not conduct heat? Or the ground state is not a QSL at all? Or there could be some other possibilities, $e.g.$, spin-lattice decoupling as suggested recently in ref.~\onlinecite{nc9_1509}? Moreover, difficulties in single crystal growth of the organic QSL candidate materials limit experimental investigations. For one of the most heavily studied QSL candidates, a kagome-lattice ZnCu$_3$(OH)$_6$Cl$_2$, the presence of 5-15\% excess Cu$^{2+}$ replacing the nonmagnetic Zn$^{2+}$ induces randomness in the magnetic exchange coupling,\cite{doi:10.1021/ja1070398} complicating the explanations of the experimental observations, and causing difficulties in understanding its ground state.\cite{nature492_406,science350_655}

For geometrically-frustrated materials, in the presence of disorder, the spin-glass phase can often emerge, since disorder and frustration are the two major ingredients for a spin glass.\cite{RevModPhys.58.801} Spins in a spin glass remain liquid like above the freezing temperature and establish short-range order below it. A spin glass is different from a QSL but can mimic it in many aspects, for example, the absence of long-range magnetic order, and moreover, as demonstrated in ref.~\onlinecite{PhysRevLett.120.087201}, the presence of continuous broad INS spectra. Below, we will give an example to elaborate this point further.

\begin{figure*}[htb]
\centerline{\includegraphics[width=5.5in]{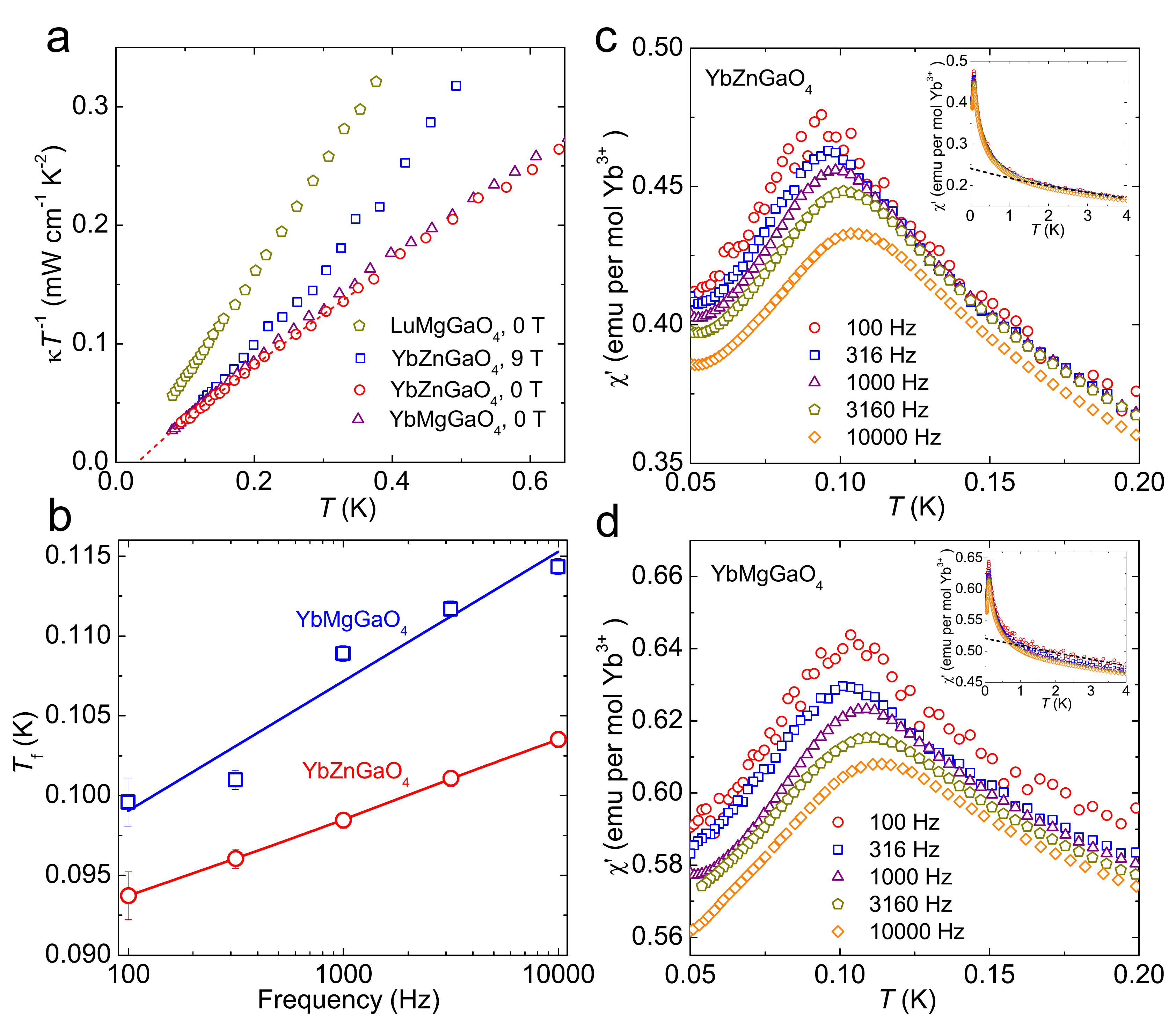}}
\caption{
{\bf Ultralow-temperature thermal conductivity and a.c. susceptibility of \ymgo and \yzgo.}
{\bf a}, Thermal conductivity results on \yzgo under zero and 9-T magnetic fields applied parallel to the $c$ axis. The dashed line is a fit to the data. For comparison, results on \ymgo and on the nonmagnetic reference compound LuMgGaO$_4$ at zero field are plotted together.\cite{PhysRevLett.117.267202} {\bf b}, Frequency dependence of the freezing temperature for both \yzgo and \ymgo, extracted from the temperature dependence of the real part of the a.c. susceptibility ($\chi^{\prime}$) shown in {\bf c} and {\bf d}. Lines through data are guides to the eye. In the insets of {\bf c} and {\bf d}, $\chi^{\prime}$ in an extended temperature range up to 4~K are plotted. Dashed lines indicate the Curie-Weiss fits for the 100-Hz data. Reprinted with permission from ref.~\onlinecite{PhysRevLett.120.087201}.
\label{fig5}}
\end{figure*}

Recently, there has been accumulating evidence that \ymgo is a promising candidate as a gapless QSL.\cite{sr5_16419,prl115_167203} By replacing Mg with Zn, Ma \et have grown a compound isostrcutural to \ymgo---\yzgo.\cite{PhysRevLett.120.087201} They have measured the new compound \yzgo utilizing various techniques, including d.c. susceptibility, specific heat, and INS. These measurements show \ymgo and \yzgo to be quite similar, both exhibiting no long-range magnetic order but prominent broad continuous gapless excitation spectra. These observations are taken as strong evidence for the gapless QSL phase in \ymgo.\cite{sr5_16419,prl115_167203,np13_117,nature540_559} By a closer look, they have found that the magnetic exchange coupling strengths in these materials are rather weak, only in the order of 0.1~meV,\cite{sr5_16419,prl115_167203,PhysRevLett.120.087201} while the disorder effect is rather strong due to the random mixing of Mg$^{2+}$/Zn$^{2+}$ and Ga$^{3+}$.\cite{PhysRevLett.118.107202,PhysRevX.8.031001}
Considering these, they have introduced disorder into a stripe-order phase, which is suggested to be the ground state for \ymgo in the absence of disorder,\cite{PhysRevLett.119.157201,PhysRevB.95.165110} and reproduced the broad excitation continua in agreement with those observed experimentally.\cite{PhysRevLett.120.087201} These results show that an antiferromagnet with disorder can also exhibit the continuum-like INS spectra.

\begin{figure}[htb]
\centerline{\includegraphics[width=0.98\linewidth]{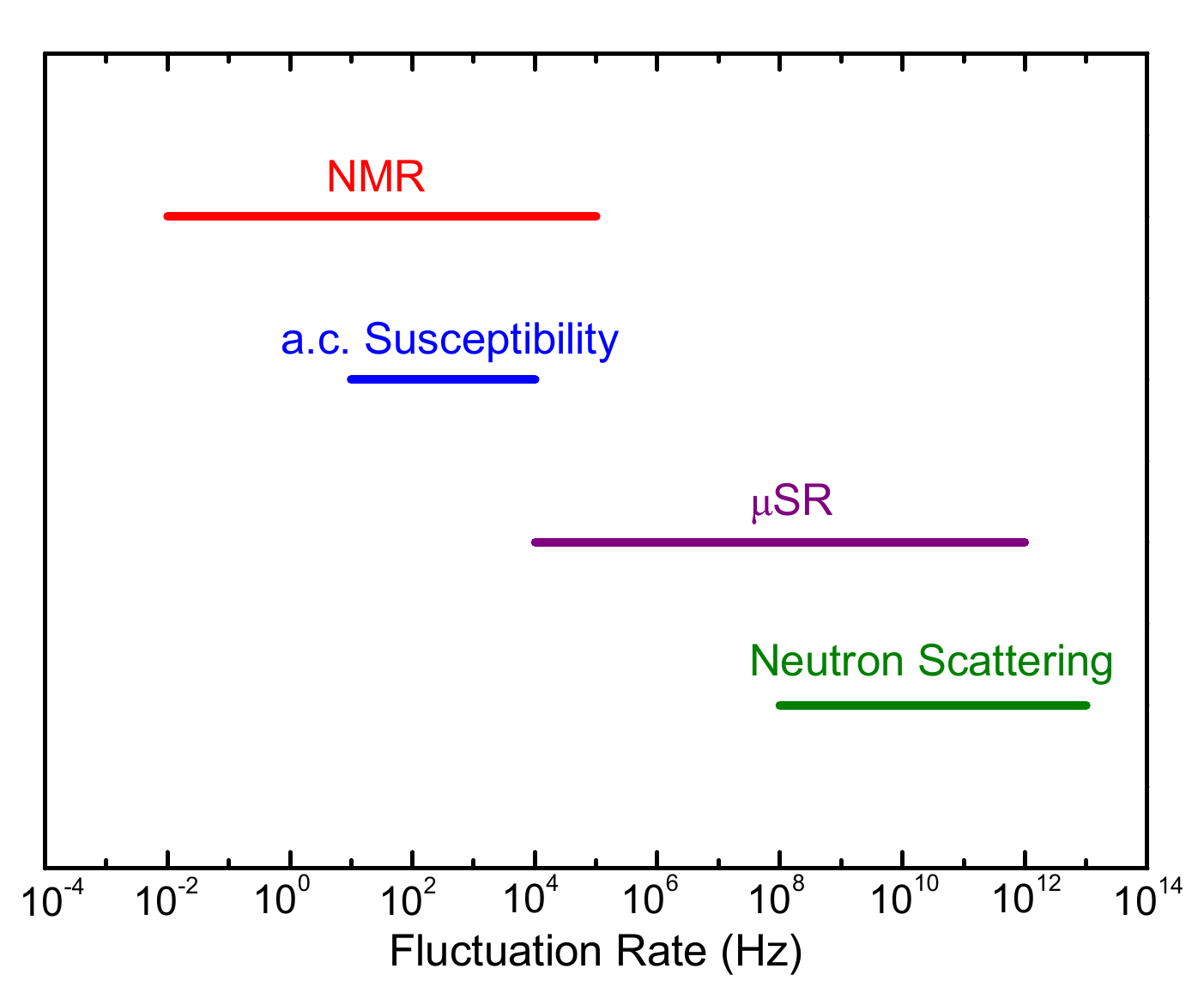}}
\caption{
{\bf A sketch of the range of the spin fluctuation rate that each technique corresponds to.} Nuclear magnetic resonance (NMR), a.c. susceptibility, muon spin relaxation ($\mu$SR), and neutron scattering, are often used probed the spin dynamics. It is very clear that different techniques are sensitive to different time scales and are complementary to each other. The a.c. susceptibility and $\mu$SR measurements are sensitive to slow and fast spin dynamics, respectively, and the overlap regime of these two techniques is very narrow. 
\label{fig6}}
\end{figure}

A key feature that distinguishes a disorder- or thermally-driven spin liquid from a QSL is whether there exist fractional spin excitations, such as the charge-neutral spin $S=1/2$ spinons in this case. For a U(1) gapless QSL as proposed for \ymgo,\cite{nature540_559} the spinons should contribute to the thermal conductivity significantly.\cite{Yamashita1246} However, as shown in Fig.~\ref{fig5}a, the thermal conductivity of \ymgo and \yzgo are almost identical, with no contribution from magnetic excitations.\cite{PhysRevLett.120.087201} Instead, it appears that the role of the gapless magnetic excitations is to scatter off phonons, causing a reduction in the thermal conductivity compared to that of the nonmagnetic sample, LuMgGaO$_4$. The absence of magnetic thermal conductivity in \ymgo and \yzgo is difficult to be reconciled with a gapless QSL state. Besides, for the U(1) QSL state proposed in \ymgo, the magnetic excitation spectra should exhibit stronger intensity at the Brillouin zone corner, in contrast to the experimental observations.\cite{np13_117,nature540_559,PhysRevLett.120.087201} Alternatively, Ma \et suggest that these materials are spin glasses, with frozen short-range correlations below the freezing temperature.\cite{PhysRevLett.120.087201} For both samples, they have identified such a phase from the a.c. susceptibility measurements, which show strong frequency-dependent peaks around 0.1~K (Fig.~\ref{fig5}b-c), characteristic of a spin glass.\cite{RevModPhys.58.801} These results demonstrate that a geometrical frustration and disorder induced spin-glass phase can behave like a QSL in various aspects.\cite{PhysRevLett.120.087201,PhysRevB.97.184413,PhysRevLett.119.157201} We note that a $\mu$SR study on \ymgo\cite{PhysRevLett.117.097201} shows that the spins remain fluctuating down to 0.07~K, below the freezing temperature reported in the a.c. susceptibility measurements.\cite{PhysRevLett.120.087201} This may be because these two techniques cover different time scales (Fig.~\ref{fig6}). More recently, Kimchi \et have proposed a disorder-induced valence-bond-glass state, which may also explain the behaviours in \ymgo and \yzgo.\cite{PhysRevX.8.031028}

\section{Outlook}

At present, we may still have not found an ideal QSL. But with such intensive and sustained efforts on theory, material searching, and characterizations, this field has been substantially moved. Below, we raise several points which we hope to motivate future research.

i), In material searching, we should still look for QSL candidates in materials with small spins and strong geometrical frustration in triangular, kagome, and pyrochlore lattice.\cite{nature464_199} {\new Ideally, a promising QSL candidate material should fulfill the following criteria: i) It should have large magnetic exchange interactions, so that it is robust to disorder and easier to be accessed experimentally; ii) It should have little disorder, so that the intrinsic physics is more prominent; iii) It should have minimal extra interactions, so that the ground state exhibits no static magnetic order.} Besides, now the material search can be extended to the SOC-assisted Mott insulators in the honeycomb lattice, which may possess anisotropic bond-dependent Kitaev interactions needed to form a Kitaev QSL.\cite{0953-8984-29-49-493002} In this regard, H$_3$LiIr$_2$O$_6$ fulfilling this condition has been suggested to be a Kitaev QSL very recently.\cite{nature554_341} Currently, the search for Kitaev QSLs have been focused on iridates and \rucl, where the electronic configuration of Ir$^{4+}$ and Ru$^{3+}$ is $d^5$ with an effective spin $j_{\rm eff}=1/2$.\cite{0953-8984-29-49-493002} Recently, Kitaev physics has also been discussed in ions which have the $d^7$ configuration, such as Co$^{2+}$, where the combination of $S=3/2$ and orbital moment $L_{\rm eff}=-1$ can also give rise to a $j_{\rm eff}=1/2$ state.\cite{PhysRevB.97.014407} Moreover, the possibility of realizing QSL physics in a square lattice featuring competing magnetic interactions (for example, competing nearest- and next-nearest-neighbour exchange couplings) has also been widely discussed in recent years.\cite{PhysRevLett.105.247001,np11_62,nc9_1085} In some occasions, there may be some other interactions that disturb the QSL state, making the material not an ideal QSL but proximate to it. Such studies are also encouraging as they not only help understand the physics of QSLs, but also offer the possibility of finding a true QSL state by suppressing the extra interactions.\cite{0953-8984-29-49-493002,ZhenMa:106101}   

ii), Identifying a QSL experimentally is a great challenge. At present, the most feasible route is to study the fractional spin excitations. Whether there exist such excitations determines whether the material is a QSL or not, and what behaviours these excitations have determines the type of QSL it can be classified into. As discussed above, there has already been some progress made in this respect. We believe that by finding more QSL candidates, improving the sample quality, and increasing the instrumental resolutions, the research on QSLs can be continuously advanced.

iii), In QSLs, orbital, spin, and lattice are quite often involved and interacting. Therefore, quantum manipulating by various tuning parameters such as magnetic field, pressure, and doping, can lead to emergent intriguing physics. For example, according to the theoretical proposals, QSLs are the parent states of high-temperature superconductors.\cite{anderson1} In this aspect, there have already been some successes in inducing superconductivity in some QSL candidates by applying pressures onto organic compounds.\cite{0034-4885-74-5-056501}
On the other hand, using a more common approach---carrier doping to induce high-temperature superconductivity, turns out to be not so successful so far.\cite{PhysRevX.6.041007} Is it the problem of the target material? For example, it is not a real QSL? Then, if a real QSL exists, will doping it with carriers eventually lead to high-temperature superconductivity as
predicted?\cite{anderson1} We note that recently, there are advances in promoting the doping capability using electric-field gating.\cite{doi:10.7566/JPSJ.83.032001} Will this technique help achieve the goal?

\medskip
\noindent {\bf Acknowledgements}\\
\noindent The work was supported by National Key Projects for Research and Development of China (Grants No.~2016YFA0300401, 2016YFA0300503, and
2016YFA0300504), the National Natural Science Foundation of China (Grants No.~11822405, 11674157, 11674158, and 11774152), the Natural Science Foundation of Jiangsu Province with Grant No.~BK20180006, Fundamental Research Funds for the Central Universities with Grant No.~020414380105, and the Office of International Cooperation and Exchanges of Nanjing University. We thank for the stimulating discussions with Yinchen He, Yi Zhou, Jia-Wei Mei, Yang Qi, Ziyang Meng, and Itamar Kimchi. 

\medskip
\noindent {\bf Correspondence}\\
\noindent Correspondence should be addressed to J.S.W.~(email: jwen@nju.edu.cn).

\medskip
\noindent {\bf Competing Interests}\\
\noindent The authors declare no competing interests.

\medskip
\noindent {\bf Author contributions}\\
J.S.W., S.L.Y., S.Y.L., W.Q.Y., and J.X.L. wrote the manuscript together.


\end{document}